\documentclass[10pt]{article}
\usepackage[T1]{fontenc}
\usepackage{geometry}
\usepackage{graphics}
\usepackage{setspace}
\makeatletter

\newcommand{\LyX}{L\kern-.1667em\lower.25em\hbox{Y}\kern-.125emX\@}

\makeatother

\begin{document}

{\par\centering {\huge An angular correlation theory for double photoionization
in a rare gas atom}\huge \par}
\vspace{0.5in}

{\par\centering {\large Dipankar Chattarji and Chiranjib Sur}\large \par}

{\par\centering \emph{Department of Physics, Visva Bharati, Santiniketan 731
235, INDIA}\par}

\vspace{0.3in}
{\small We consider the process of double photoionization (DPI) in a rare gas
atom as a two-step process, namely (i) photoionization in an inner shell followed
by (ii) the emission of an Auger electron from an outer shell. An angular correlation
function for the two emitted electrons is defined by analogy with the theory
of angular correlation in nuclear physics. An expression is obtained for this
angular correlation function by a statistical method which makes use of the density and efficiency
operators. The latter takes care of the attenuation of the probability of detection of an electrons due to the geometrical properties of the detector.
 Theoretical
values of the angular correlation function are obtained for DPI in xenon and
these are shown to be in good agreement with the experimental results given
by K\( \ddot{a} \)mmerling and Schmidt {[}\ref{schmidt}{]}.{\small \par}

\vspace{0.2in}
\vspace{0.2in}

\section{Introduction}

The theory of angular correlation was originally formulated in nuclear physics.
This was done for two possible kinds of phyical situation.

(i) Angular correlation of successive nuclear radiations emitted by a radioactive
nuclear species, e.g. a \( \gamma -\gamma  \) cascade. 

(ii) Angular correlation of successive nuclear radiations resulting from the
more general case of a nuclear scattering process. 

The earliest paper on this subject was written by Hamilton {[}\ref{hamilton}{]}
for the case of a \( \gamma -\gamma  \) cascade. Later on, Gardner {[}\ref{gardner}{]}
wrote down the angular correlation function using the wave functions of the
states occurring in a cascade decay. Racah {[}\ref{racah}{]} obtained thereafter
a simple expression for the angular correlation function in terms of the angle between
the directions of emission of two successive radiations emitted by a nucleus.
He did not worry about the history of this nucleus, and his theory held regardless
of whether the emitted radiations consisted of bosons or fermions. Subsequently,
Biedenharn and Rose {[}\ref{biedenharn}{]} extended Racah's work to give a
general form of the angular correlation function for successive nuclear radiations.
About the same time, Coester and Jauch {[}\ref{coester}{]} gave a theory in
the more general context of nuclear reactions. Their paper gives a derivation
of the angular correlation function which brings out the statistical nature
of the problem. Later, Devons and Goldfarb {[}\ref{goldfarb}{]} wrote a detailed
review of angular correlations up to that point.

In our problem a rare gas atom absorbs a photon having specified properties i.e.
energy, spin etc. As a result, the atom emits a photo-electron giving
rise to a singly charged ion. This ion now de-excites by emitting an Auger electron.
This yields a doubly charged residual ion. \\
Note two basic features of this problem.

(i) The initial atomic state is randomly oriented ( an improper state or ``Gemisch'').

(ii) Only limited information is available about the emitted electrons, usually
their directions of motion. 

The density matrix or statistical matrix \( \rho  \) was introduced into quantum
mechanics to provide for the discussion of just such a situation {[}\ref{blum},
\ref{fano}{]}. It forms an important ingredient of the nuclear theory of angular
correlations. It will naturally figure in our atomic theory of angular correlations as well. 

In section 2 we shall try to construct an angular correlation function for the
atomic problem outlined above. This is done in section 2.1 using a statistical approach
. In section 3 we report our calculation for the xenon atom which
was experimentally studied by Schmidt and his co-workers{[}\ref{schmidt}{]}.

\section{An angular correlation theory for atoms : the double photoionization problem}

Double photoionization (DPI) occurs when an atomic target consisting of rare
gas atoms is irradiated with photons from an advanced light source, e.g. a synchrotron,
and an atom emits two electrons in quick succession. In case the time interval
between the emission of these two electrons is very small, one could think of
DPI as a one-step process. A considerable amount of work has already been done
on this {[}\ref{selles1},\ref{selles2}{]}. However, in case the time interval
between the successive emission of the two electrons is substantially longer
than the time taken by the photo-electron to leave the interaction zone, DPI
may be regarded as a two-step process {[}\ref{we}{]}. This in its turn will
depend on the energy imparted to the atomic system by the incident photon.

We shall now try to construct a theory for the two-step process described above,
keeping in view the earlier work done in nuclear physics. Since this is a two-step
process mediated by electromagnetic interaction at two vertices, we expect it
to be a second order process as compared to normal photoionization (PI). Hence
the probability of its occurrence will be much lower than that of PI, and its
detection will call for much greater precision.

Consider a randomly oriented rare gas atom in a \( ^{1}S^{e} \) state. It absorbs
a photon and after a certain time interval emits a photo-electron from an inner
shell, giving a single-vacancy ionic state. This intermediate ionic state now
de-excites by emitting an Auger electron, typically from an outer shell, giving
rise to a two-vacancy final state {[}\ref{we}{]}. \\
So our process amounts to 

\begin{equation}
\label{1}
h\nu +\mathbf{A}\longrightarrow \mathbf{A}^{+}+e_{1}^{-}\longrightarrow \mathbf{A}^{++}+e_{1}^{-}+e_{2}^{-}\, .
\end{equation}

The initial state of our system is characterized by the set of quantum
numbers \( (J_{a}M_{a}\alpha_{a})\), or by virtual quantum numbers \((J^{\prime }_{a}M^{\prime }_{a}\alpha^{\prime}_{a}) \) arising from possible interaction with other atoms and electrons.
Here \((J_{a}M_{a})\) or \((J^{\prime}_{a}M^{\prime}_{a})\) are angular momentum quantum numbers, and \(\alpha_{a},\alpha^{\prime}_{a}\) denote the set of remaining quantum numbers characterizing the system.

We wish to obtain an angular correlation function for the two emitted electrons
in terms of the angle \( \theta  \) between their directions of emission.

\subsection{The method of efficiency and density operators}

We now approach the problem of angular correlation from a statistical standpoint.
We note that there is a certain probability for the atom to attain the final
state given above. This is described by the appropriate matrix element of the
density or statistical operator \( \rho  \) as defined in the literature {[}\ref{blum},\ref{terharr}{]}.
Now, even if the atom goes over to the final state, because of the finite size
of our detecting equipment and other limiting factors, this event may or may
not be detected. There is thus a certain probability \( \epsilon \, (0\leq \epsilon \leq 1) \)
that the event will be detected. This probability is represented by the efficiency
operator \( \varepsilon  \) which will depend on the size, position and 
geometrical arrangement of the detecting equipment.

Now, how does the angular correlation function relate to the operators \( \rho  \)
and \( \varepsilon  \)? We shall try to answer this question in the following
manner. We start out by defining

\begin{equation}
\label{2}
\begin{array}{cc}
\overline{\varepsilon } & =\sum _{Q}\varepsilon _{Q}\left\langle Q\right| \rho \left| Q\right\rangle \\
 & =\sum _{Q}\varepsilon _{Q}\rho _{QQ}\\
 & =Tr(\varepsilon \rho )\, .
\end{array}
\end{equation}
 Here \( \varepsilon _{Q} \) is the efficiency or probability of detection
of the state described by quantum numbers \( Q \), and \( \rho _{QQ} \) the
probability of the system being in the particular state \( Q \). 

From the elements of statistical mechanics, we know that \( \overline{\varepsilon } \)
is the expectation value ( or average value) of the efficiency operator \( \varepsilon  \)
{[}\ref{terharr}{]}. We shall presently see that the angular correlation function
is simply related to \( \overline{\varepsilon } \).

Since \( \overline{\varepsilon } \) is the trace of a matrix 
it is invariant under a unitary transformation in Hilbert space.

We now ask the question : what are the transformation properties of the matrices
\( \varepsilon  \) and \( \rho  \) as we go from one unitary representation
to another? It is easily seen that \( \varepsilon  \) and \( \rho  \) are
tensor operators. 
Hence they are also
called the efficiency and density (or statistical) tensors respectively {[}\ref{coester}{]}.
The fact that \( \varepsilon  \) and \( \rho  \) are both tensor operators
makes them amenable to further analysis.

Since the initial atomic system is randomly oriented, we have rotational symmetry
and angular momentum is conserved. Hence our state vectors are eigenvectors
of \( \mathbf{J}^{2} \)and \( {J_{z}} \) . In such
a case the matrix elements of a tensor operator have a simple geometric dependence
on the magnetic quantum numbers. This is given by the 
Wigner-Eckart theorem.

Using this theorem we write the matrix element of the density operator as

\begin{equation}
\label{3}
\left\langle J_{a}M_{a}\alpha _{a}\right| \rho \left| J^{\prime }_{a}M^{\prime }_{a}\alpha _{a}^{\prime }\right\rangle =\sum _{k_{a}\kappa _{a}}(-1)^{J^{\prime }_{a}-M^{\prime }_{a}}C_{M_{a}M^{\prime }_{a}\kappa _{a}}^{J_{a}J^{\prime }_{a}k_{a}}\rho _{k_{a}\kappa _{a}}(J_{a}\alpha _{a},J^{\prime }_{a}\alpha _{a}^{\prime }),
\end{equation}
 where \( \mathbf{k_{a}}=\mathbf{J_{a}}+\mathbf{J_{a}}^{\prime } \)and
\( \kappa _{a} \) is the projection of \( k_{a} \). \\
Similarly, the matrix element of the efficiency operator is given by

\begin{equation}
\label{4}
\left\langle J_{a}M_{a}\alpha _{a}\right| \varepsilon \left| J^{\prime }_{a}M^{\prime }_{a}\alpha _{a}^{\prime }\right\rangle =\sum _{k_{a}\kappa _{a}}(-1)^{J^{\prime }_{a}-M^{\prime }_{a}}C_{M_{a}M^{\prime }_{a}\kappa _{a}}^{J_{a}J^{\prime }_{a}k_{a}}\varepsilon _{k_{a}\kappa _{a}}(J_{a}\alpha _{a},J^{\prime }_{a}\alpha _{a}^{\prime }).
\end{equation}

Using the unitarity property of Clebsch-Gordan coefficients we get the density
tensor of rank \( k_{a} \) with \( (2k_{a}+1) \) components

\begin{equation}
\label{6}
\rho _{k_{a}\kappa _{a}}(J_{a}\alpha _{a},J^{\prime }_{a}\alpha _{a}^{\prime })=\sum _{M_{a}M^{\prime }_{a}}(-1)^{M_{a}-M^{\prime }_{a}}C_{M_{a}M^{\prime }_{a}\kappa _{a}}^{J_{a}J^{\prime }_{a}k_{a}}\left\langle J_{a}M_{a}\alpha _{a}\right| \rho \left| J^{\prime }_{a}M^{\prime }_{a}\alpha _{a}^{\prime }\right\rangle .
\end{equation}
 A similar expression can be obtained for the components of the efficiency tensor.

So the expectation value of \( \varepsilon  \) becomes 

\begin{equation}
\label{5}
\overline{\varepsilon }=Tr(\varepsilon \rho )=\sum _{J_{a}J^{\prime }_{a}\alpha _{a}\alpha _{a}^{\prime }k_{a}\kappa _{a}}\rho _{k_{a}\kappa _{a}}(J_{a}\alpha _{a},J^{\prime }_{a}\alpha _{a}^{\prime })\varepsilon ^{\star }_{k_{a}\kappa _{a}}(J_{a}\alpha _{a},J^{\prime }_{a}\alpha _{a}^{\prime }).
\end{equation}

Our choice of phase {[}\ref{edmonds}{]} ensures that \( \rho _{k_{a}\kappa _{a}} \)
is a Hermitian tensor. It satisfies the relation 

\begin{equation}
\label{7}
\rho ^{\star }_{k_{a}\kappa _{a}}(J_{a}\alpha _{a},J^{\prime }_{a}\alpha _{a}^{\prime })=(-1)^{J^{\prime }_{a}-J_{a}+\kappa _{a}}\rho _{k_{a}-\kappa _{a}}(J^{\prime }_{a}\alpha _{a}^{\prime },J_{a}\alpha _{a}).
\end{equation}
 Some simplification yields the result

\begin{equation}
\label{8}
\rho _{k_{a}\kappa _{a}}(J_{a}\alpha _{a},J^{\prime }_{a}\alpha _{a}^{\prime })=\frac{\delta _{k_{a}0}\delta _{\kappa _{a}0}\delta _{J_{a}J^{\prime }_{a}}}{\sqrt{2J_{a}+1}}\left\langle J_{c}\right\Vert j_{2}\left\Vert J_{b}\right\rangle \left\langle J_{c}\right\Vert j^{\prime }_{2}\left\Vert J_{b}\right\rangle ^{\star }\left\langle J_{b}\right\Vert j_{1}\left\Vert J_{a}\right\rangle \left\langle J_{b}\right\Vert j^{\prime }_{1}\left\Vert J_{a}\right\rangle ^{\star }.
\end{equation}

The first decay, namely the photoionization process, is characterized by the
relation

\begin{equation}
\label{9}
\mathbf{J_{a}}=\mathbf{J_{b}}+\mathbf{j_{1}}.
\end{equation}
 Since the detection of the photo-electron and that of the singly charged ion
are independent events, the joint probability of their detection is given by
the product of the individual probabilities. Hence we can write

\begin{equation}
\label{10}
\begin{array}{cc}
\left\langle J_{a}M_{a}\alpha _{a}\right| \varepsilon \left| J^{\prime }_{a}M^{\prime }_{a}\alpha _{a}^{\prime }\right\rangle  & =\sum \left\langle J_{b}M_{b}\alpha _{b}\right| \varepsilon \left| J^{\prime }_{b}M^{\prime }_{b}\alpha _{b}^{\prime }\right\rangle \left\langle j_{1}\mu _{1}\right| \varepsilon \left| j^{\prime }_{1}\mu ^{\prime }_{1}\right\rangle \\
 & \times C_{M_{b}\mu _{1}M_{a}}^{\star J_{b}j_{1}J_{a}}C_{M^{\prime }_{b}\mu ^{\prime }_{1}M^{\prime }_{a}}^{J_{b}j^{\prime }_{1}J^{\prime }_{a}}\, .
\end{array}
\end{equation}
 Here \( (J_{b}M_{b}) \), \( (J^{\prime }_{b}M^{\prime }_{b}) \) describe
 the intermediate state and \( \alpha _{b},\alpha _{b}^{\prime } \)
denote the remaining quantum numbers. The summation in Eq. (\ref{10}) extends
over \( M_{b}M^{\prime }_{b}\mu_{1} \mu ^{\prime }_{1} \).

Then

\begin{equation}
\label{11}
\begin{array}{ccc}
\varepsilon _{k_{a}\kappa _{a}}(J_{a}\alpha _{a},J^{\prime }_{a}\alpha _{a}^{\prime }) & = & \sum_{k_{1}\kappa_{1}k_{b}\kappa_{b}} \varepsilon _{k_{b}\kappa _{b}}(J_{b}\alpha _{b},J^{\prime }_{b}\alpha _{b}^{\prime })\varepsilon _{k_{1}\kappa _{1}}(j_{1}j^{\prime }_{1})\\
 &  & \times C_{\kappa _{b}\kappa _{1}\kappa _{a}}^{\star k_{b}k_{1}k_{a}}\sqrt{2J_{a}+1}\sqrt{2J^{\prime}_{a}+1}\sqrt{2k_{b}+1}\sqrt{2k_{1}+1}\left\{ \begin{array}{ccc}
J_{b} & j_{1} & J_{a}\\
J^{\prime }_{b} & j^{\prime }_{1} & J^{\prime }_{a}\\
k_{b} & k_{1} & k_{a}
\end{array}\right\} \, ,
\end{array}
\end{equation}
 with the relations

\( \mathbf{J_{a}}=\mathbf{J_{b}}+\mathbf{j_{1}} \)
, \( \mathbf{k_{a}}=\mathbf{J_{a}}+\mathbf{J_{a}}^{\prime } \)
, \( \mathbf{k_{b}}=\mathbf{J_{b}}+\mathbf{J_{b}}^{\prime } \)
and \( \mathbf{k_{1}}=\mathbf{j_{1}}+\mathbf{j_{1}}^{\prime } \).
\\
Since the intermediate singly ionized state decays into a residual doubly ionized
atom and an Auger electron, we can factorise the efficiency operator of the singly
ionized atom in terms of the efficiency operators of the residual doubly ionized
atom and the Auger electron. 

Since the residual doubly ionized atom is left in a sharp eigenstate
with the quantum number \( J_{c} \) and no further measurement is made on
it, we get the efficiency operator

\begin{equation}
\label{17a}
\varepsilon _{k_{c}\kappa _{c}}(J_{c}\alpha _{c},J^{\prime }_{c}\alpha _{c}^{\prime })=\sqrt{2J_{c}+1}\delta _{k_{c}0}\delta _{\kappa _{c}0}\delta _{J_{c}J^{\prime }_{c}}\delta _{\alpha _{c}\alpha ^{\prime }_{c}}.
\end{equation}

A matrix element of the efficiency operator for each electron has the form

\begin{equation}
\label{12}
\begin{array}{cc}
\varepsilon _{k_{i}\kappa _{i}}(j_{i}j^{\prime }_{i}) & =\sum _{\mu _{i}\mu _{i}\prime }(-1)^{j_{i}-j^{\prime }_{i}}C_{\mu _{i}-\mu ^{\prime }_{i}\kappa _{i}}^{j_{i}j^{\prime }_{i}k_{i}}\left\langle j_{i}\mu _{i}\right| \varepsilon \left| j^{\prime }_{i}\mu ^{\prime }_{i}\right\rangle \\
 & =\sum _{\mu _{i}\mu ^{\prime }_{i}\sigma _{i}\sigma ^{\prime }_{i}}(-1)^{j^{\prime }_{i}-\mu ^{\prime }_{i}}C_{\mu _{i}-\mu ^{\prime }_{i}\kappa _{i}}^{j_{i}j^{\prime }_{i}k_{i}}\left\langle j_{i}\mu _{i}\right. \left| \Omega _{i}\sigma _{i}\right\rangle \\
 & \times \left\langle \sigma _{i}\right| \varepsilon \left| \sigma ^{\prime }_{i}\right\rangle \left\langle \Omega _{i}\sigma ^{\prime }_{i}\right. \left| j^{\prime }_{i}\mu ^{\prime }_{i}\right\rangle ,
\end{array}
\end{equation}
 with \( i=1,2 \). \( i=1 \) denotes the photo-electron and \( i=2 \) the
Auger electron. \\
Using the axis of the detector as the quantization axis for each detected electron, we get

\begin{equation}
\label{13}
\left\langle \Omega _{i}\sigma _{i}\right. \left| j_{i}\mu _{i}\right\rangle =\sum_{\kappa_{i}} \left\langle 0\sigma _{i}\right. \left| j_{i}\kappa _{i}\right\rangle D_{\mu _{i}\kappa _{i}}^{j_{i}\star }(\Re _{i}).
\end{equation}
 \( D(\Re _{i}) \) is the corresponding rotation matrix for the \( i\)th electron.

Then Eq. (\ref{12}) becomes
\begin{equation}
\label{16}
\varepsilon _{k_{i}\kappa_{i} }(j_{i}j_{i}^{\prime })=\sum _{\kappa_{i} ^{\prime }}c_{k_{i}\kappa_{i} ^{\prime }}(j_{i}j_{i}^{\prime })D_{\kappa_{i} \kappa_{i} ^{\prime }}^{k_{i}\star }(\Re_{i} ),
\end{equation}
 where

\begin{equation}
\label{15}
\begin{array}{cc}
c_{k_{i}\kappa _{i}}(j_{i}j^{\prime }_{i})= & \Im \sum _{\mu _{i}\mu \prime _{i}}(-1)^{j^{\prime }_{i}-\mu ^{\prime }_{i}}\left\langle 0\sigma _{i}\right. \left| j_{i}\mu _{i}\right\rangle ^{\star }\\
 & \times \left\langle 0\sigma ^{\prime }_{i}\right. \left| j^{\prime }_{i}\mu ^{\prime }_{i}\right\rangle C_{\mu _{i}-\mu ^{\prime }_{i}\kappa _{i}}^{j_{i}j^{\prime }_{i}k_{i}}\times \left\langle \sigma _{i}\right| \varepsilon \left| \sigma ^{\prime }_{i}\right\rangle .
\end{array}
\end{equation}
 The symbol \( \Im  \) represents a summation over the spin of the emitted
electrons and depends on the characteristics of the detector. We now take
the summation \( \Im f(\sigma \sigma ^{\prime }) \) to be equivalent to \( \sum \left\langle \sigma \right| \varepsilon \left| \sigma ^{\prime }\right\rangle f(\sigma \sigma ^{\prime }) \)
. 

Making use of properties of Clebsch-Gordan coefficients and \( 9-j \) symbols,
and also using the additivity of rotation matrices {[}\ref{edmonds}{]} 

\begin{equation}
\label{19}
\begin{array}{ccc}
\sum _{\nu }D_{\nu \kappa _{1}}^{k}(\Re _{1})D_{\nu \kappa _{2}}^{k\star }(\Re _{2}) & = & \sum _{\nu }D_{\nu \kappa _{1}}^{k}(\Re _{1})D_{\kappa _{2}\nu }^{k}(\Re ^{-1}_{2})\\
 & = & D_{\kappa _{2}\kappa _{1}}^{k}(\Re ^{-1}_{2}\Re _{1}),
\end{array}
\end{equation}
 we get

\begin{equation}
\label{18}
\begin{array}{ccc}
\overline{\varepsilon } & = & (2J_{b}+1)(-)^{J_{a}+J_{c}-2J_{b}}\sum (-1)^{k-j_{1}-j_{2}}w(J_{b}J^{\prime }_{b}j_{1}j^{\prime }_{1},kJ_{a})\\
 &  & \times w(J_{b}J^{\prime }_{b}j_{2}j^{\prime }_{2},kJ_{c})c_{k\kappa _{1}}(j^{\prime }_{1}j_{1})c^{\star }_{k\kappa _{2}}(j_{2}j^{\prime }_{2})D_{\kappa _{1}\kappa _{2}}^{k}(\theta _{1}\theta _{2}\theta _{3})\\
 &  & \times \left\langle J_{c}\right\Vert j_{2}\left\Vert J_{b}\right\rangle \left\langle J_{c}\right\Vert j^{\prime }_{2}\left\Vert J_{b}\right\rangle ^{\star }\left\langle J_{b}\right\Vert j_{1}\left\Vert J_{a}\right\rangle \left\langle J_{b}\right\Vert j^{\prime }_{1}\left\Vert J_{a}\right\rangle ^{\star }.
\end{array}
\end{equation}
 In Eq. (\ref{18}) the summation is over \( j_{1},j^{\prime }_{1},j_{2},j^{\prime }_{2},k,\kappa _{1}\, and\, \kappa _{2} \)
. 

If the electrons are unpolarized, 
or if the detectors are insensitive to polarization, \( \kappa _{1}=\kappa _{2}=0 \)
and \( D_{00}^{k}(\theta _{1}\theta _{2}\theta _{3})=P_{k}(cos\theta _{2}) \)
. We now identify the angle \( \theta _{2} \) with \( \theta  \), the angle
between the directions of emission of the photo-electron and the Auger electron.\\
From the Hermitian character of the efficiency and density matrices, it
follows that

\begin{equation}
\label{21}
c_{k\kappa }(jj^{\prime })=(-1)^{\kappa }c^{\star }_{k\kappa }(j^{\prime }j).
\end{equation}
 For our case \( \kappa _{1}=\kappa _{2}=0 \), and 

\begin{equation}
\label{22}
c_{k0}(jj^{\prime })=\frac{\sqrt{2j+1}\sqrt{2j{\prime }+1}}{4\pi }(-1)^{j-\frac{1}{2}+k}C^{jj^{\prime }k}_{\frac{1}{2}-\frac{1}{2}0}\, .
\end{equation}
 We can thus write,

\begin{equation}
\label{23}
\begin{array}{ccc}
\overline{\varepsilon } & = & (-1)^{J_{a}+J_{c}-2J_{b}}\sqrt{2J_{b}+1}\sum_{k} (-1)^{k-j_{1}-j_{2}}w(J_{b}J^{\prime }_{b}j_{1}j^{\prime }_{1},kJ_{a})\\
 &  & \times w(J_{b}J^{\prime }_{b}j_{2}j^{\prime }_{2},kJ_{c})\left\langle J_{c}\right\Vert j_{2}\left\Vert J_{b}\right\rangle \left\langle J_{c}\right\Vert j^{\prime }_{2}\left\Vert J_{b}\right\rangle ^{\star }\\
 &  & \times \left\langle J_{b}\right\Vert j_{1}\left\Vert J_{a}\right\rangle \left\langle J_{b}\right\Vert j^{\prime }_{1}\left\Vert J_{a}\right\rangle ^{\star }c_{k0}(j_{1}j^{\prime }_{1})c^{\star }_{k0}(j_{2}j^{\prime }_{2})P_{k}(cos\theta ).
\end{array}
\end{equation}
 Here \( k \) is an even integer ranging from \( 0 \) to \( k_{max} \), \( k_{max} \)
being defined as follows. Let \( \left\{ \left\{ j_{1}+j^{\prime }_{1}\right\} _{max},\left\{ j_{2}+j^{\prime }_{2}\right\} _{max}\right\} _{min}=p \).
Then \( k_{max}=p \) if \( p \) is even and \( k_{max}=p-1 \) if \( p \)
is odd. \\
 We now express \( \left\langle J_{c}\right\Vert j_{2}\left\Vert J_{b}\right\rangle  \)
in terms of \( \left\langle J_{b}\right\Vert j_{2}\left\Vert J_{c}\right\rangle  \)
. Though the reduced matrix elements are neither real nor Hermitian, it happens that  

\begin{equation}
\label{24}
\sqrt{2J_{b}+1}\left\langle J_{a}\right\Vert j\left\Vert J_{b}\right\rangle =(-1)^{J_{a}-j+J_{b}}\sqrt{2J_{a}+1}\left\langle J_{b}\right\Vert j\left\Vert J_{a}\right\rangle ^{\star }.
\end{equation}
 This gives

\begin{equation}
\label{25}
\begin{array}{ccc}
\overline{\varepsilon } & = & \sum _{k}(-1)^{j_{1}+j_{2}}c_{k0}(j_{1}j^{\prime }_{1})c^{\star }_{k0}(j_{2}j^{\prime }_{2})\\
 &  & \times \left\langle J_{c}\right\Vert j_{2}\left\Vert J_{b}\right\rangle \left\langle J_{c}\right\Vert j^{\prime }_{2}\left\Vert J_{b}\right\rangle ^{\star }\left\langle J_{b}\right\Vert j_{1}\left\Vert J_{a}\right\rangle \left\langle J_{b}\right\Vert j^{\prime }_{1}\left\Vert J_{a}\right\rangle ^{\star }\\
 &  & \times w(J_{b}J^{\prime }_{b}j_{1}j^{\prime }_{1},kJ_{a})w(J_{b}J^{\prime }_{b}j_{2}j^{\prime }_{2},kJ_{c})P_{k}(cos\theta ).
\end{array}
\end{equation}

If the finite size of the detector is taken into account, the efficiency of detection
described by the matrix element of the efficiency operator must be changed slightly.
Then we have to introduce \( z_{k} \) as the attenuation factor due to the finite
size of the detector. We assume the detector to be axially symmetric {[}\ref{rose},\ref{ferguson}{]}.
The efficiency tensor described by Eq.(\ref{16}) is now written as

\begin{equation}
\label{26}
\varepsilon _{k_{i}\kappa_{i} }(j_{i}j_{i}^{\prime })=\sum _{\kappa_{i} ^{\prime }}z_{k_{i}}c_{k_{i}\kappa_{i} ^{\prime }}(j_{i}j_{i}^{\prime })D_{\kappa_{i} \kappa_{i} ^{\prime }}^{k_{i}\star }(\Re_{i} ).
\end{equation}
 So the expectation value in our case becomes

\begin{equation}
\label{27}
\begin{array}{ccc}
\overline{\varepsilon } & = & \sum _{k}z_{k}(1)_{k}z_{k}(2)(-1)^{j_{1}+j_{2}}c_{k0}(j_{1}j^{\prime }_{1})c^{\star }_{k0}(j_{2}j^{\prime }_{2})\\
 &  & \times \left\langle J_{c}\right\Vert j_{2}\left\Vert J_{b}\right\rangle \left\langle J_{c}\right\Vert j^{\prime }_{2}\left\Vert J_{b}\right\rangle ^{\star }\left\langle J_{b}\right\Vert j_{1}\left\Vert J_{a}\right\rangle \left\langle J_{b}\right\Vert j^{\prime }_{1}\left\Vert J_{a}\right\rangle ^{\star }\\
 &  & \times w(J_{b}J^{\prime }_{b}j_{1}j^{\prime }_{1},kJ_{a})w(J_{b}J^{\prime }_{b}j_{2}j^{\prime }_{2},kJ_{c})P_{k}(cos\theta ).
\end{array}
\end{equation}
 Note that the \( \theta  \) dependence of \( \overline{\varepsilon } \) is
contained in the function

\begin{equation}
\label{28}
\begin{array}{cc}
W(\theta ) & =\sum _{k}z_{k}(1)_{k}z_{k}(2)(-1)^{j_{1}+j_{2}}c_{k0}(j_{1}j^{\prime }_{1})c^{\star }_{k0}(j_{2}j^{\prime }_{2})\\
 & \times w(J_{b}J^{\prime }_{b}j_{1}j^{\prime }_{1},kJ_{a})w(J_{b}J^{\prime }_{b}j_{2}j^{\prime }_{2},kJ_{c})P_{k}(cos\theta ).
\end{array}
\end{equation}

We now define \( W(\theta ) \) to be the angular correlation function for the two emitted electrons where \(\theta\) is the angular separation between their directions of emission
{[}\ref{our1}{]}. It is clear that angular correlation between the directions
of emission of the photo-electron and the Auger electron is a direct manifestation
of the efficiency of the observing equipment. In section 3 we shall see that the angular correlation
function so defined agrees closely with the measured angular correlation in
the DPI experiments on xenon {[}\ref{schmidt}{]}. This confirms that 
the observing equipment does play a role, introducing
an element of probability which finds expression in the angular correlation function. 

This definition will have to be modified if it is possible for the photo-electron
to be emitted into more than one angular momentum channels. In section 3 we
shall see how this modification can be made.

Note that Eq.(\ref{28}) holds formally not only for double photoionization in atoms, but generally for two-step angular correlation experiments involving either fermions or bosons.

\section{Calculation and results}

We consider the problem of DPI in xenon. Neutral xenon atoms are irradiated
with a photon beam of energy \( 94.5\, eV \). This leads to photo-ionization in the 
\( 4d_{5/2} \) shell followed by an \( N_{5}-O_{2,3}O_{2,3}\, ^{1}S_{0} \)
Auger decay. Using the dipole approximation the possible photoionization channels
are 
\(e) 4d_{5/2}\longrightarrow \epsilon _{p}f_{7/2} \) , \(f) 4d_{5/2}\longrightarrow \epsilon _{p}f_{5/2} \)
and \(g) 4d_{5/2}\longrightarrow \epsilon _{p}p_{3/2} \) respectively. The Auger
transition is characterized by only one partial wave \( \epsilon _{A}d_{5/2} \){[}\ref{dc}{]}.
These transitions are governed by the corresponding selection rules for photoionization
and Auger transitions.

Since the initial photoionization process is not characterized by a single angular momentum
quantum number but by angular momentum quantum numbers corresponding to three possible
channels, the angular correlation function described in section 2.1 above will
be modified. The total intensity will,
however, remain unchanged. If the photoionization channels are described by
the total angular momentum quantum numbers \( j^{e}_{1},j^{f}_{1} \) and
\( j^{g}_{1} \) and the Auger electron by \( j_{2} \), then the
expectation value of the efficiency operator becomes

\begin{equation}
\label{res-1}
\overline{\varepsilon (\theta )}=\overline{\varepsilon_{e}(\theta )}+\overline{\varepsilon_{f}(\theta )}+\overline{\varepsilon_{g}(\theta )}+\overline{\varepsilon_{ef}(\theta )}+\overline{\varepsilon_{fg}(\theta )}+\overline{\varepsilon_{ge}(\theta )\, .}
\end{equation}
 Here

\begin{equation}
\label{res-2}
\begin{array}{ccc}
\overline{\varepsilon_{e}(\theta )} & = & \sum _{k}z_{k}(1)z_{k}(2)(-1)^{j^{e}_{1}+j_{2}}c_{k0}(j^{e}_{1}j^{e}_{1})c^{\star }_{k0}(j_{2}j_{2})\\
 &  & \times \left| \left\langle J_{c}\right\Vert j_{2}\left\Vert J_{b}\right\rangle \right| ^{2}\left| \left\langle J_{a}\right\Vert j^{e}_{1}\left\Vert J_{b}\right\rangle \right| ^{2}\\
 &  & \times w(J_{b}J_{b}j^{e}_{1}j^{e}_{1},kJ_{a})w(J_{b}J_{b}j_{2}j_{2},kJ_{c})P_{k}(cos\theta ).
\end{array}
\end{equation}
The expectation values \( \overline{\varepsilon_{f}(\theta )} \) and \( \overline{\varepsilon_{g}(\theta )} \)
have the same form with \( j^{e}_{1}\longrightarrow j^{f}_{1} \) and \( j^{e}_{1}\longrightarrow j^{g}_{1} \)
respectively.    The quantity \( \overline{\varepsilon _{ij}(\theta )} \) is an interference
term arising from interaction between photoionization channels \( i \) and
\( j \) \( (i,j=e,f,g\, with\, i\neq j) \) . 

\begin{equation}
\label{res-3}
\begin{array}{ccc}
\overline{\varepsilon_{ef}(\theta )} & = & \sum _{k}z_{k}(1)z_{k}(2)(-1)^{j_{2}}\left| \left\langle J_{c}\right\Vert j_{2}\left\Vert J_{b}\right\rangle \right| ^{2}\left\langle J_{a}\right\Vert j^{e}_{1}\left\Vert J_{b}\right\rangle \left\langle J_{a}\right\Vert j^{f}_{1}\left\Vert J_{b}\right\rangle \\
 &  & \times [(-1)^{j^{e}_{1}}c_{k0}(j^{e}_{1}j^{f}_{1})+(-1)^{j^{f}_{1}}c_{k0}(j^{f}_{1}j^{e}_{1})]c^{\star }_{k0}(j_{2}j_{2})\\
 &  & \times w(J_{b}J_{b}j^{e}_{1}j^{f}_{1},kJ_{a})w(J_{b}J_{b}j_{2}j_{2},kJ_{c})P_{k}(cos\theta ),
\end{array}
\end{equation}

\begin{equation}
\label{res-4}
\begin{array}{ccc}
\overline{\varepsilon_{fg}(\theta )} & = & \sum _{k}z_{k}(1)z_{k}(2)(-)^{j_{2}}\left| \left\langle J_{c}\right\Vert j_{2}\left\Vert J_{b}\right\rangle \right| ^{2}\left\langle J_{a}\right\Vert j^{f}_{1}\left\Vert J_{b}\right\rangle \left\langle J_{a}\right\Vert j^{g}_{1}\left\Vert J_{b}\right\rangle \\
 &  & \times [(-1)^{j^{f}_{1}}c_{k0}(j^{f}_{1}j^{g}_{1})+(-1)^{j^{g}_{1}}c_{k0}(j^{g}_{1}j^{f}_{1})]c^{\star }_{k0}(j_{2}j_{2})\\
 &  & \times w(J_{b}J_{b}j^{g}_{1}j^{f}_{1},kJ_{a})w(J_{b}J_{b}j_{2}j_{2},kJ_{c})P_{k}(cos\theta )
\end{array}
\end{equation}
 and

\begin{equation}
\label{res-5}
\begin{array}{ccc}
\overline{\varepsilon_{ge}(\theta )} & = & \sum _{k}z_{k}(1)z_{k}(2)(-1)^{j_{2}}\left| \left\langle J_{c}\right\Vert j_{2}\left\Vert J_{b}\right\rangle \right| ^{2}\left\langle J_{a}\right\Vert j^{g}_{1}\left\Vert J_{b}\right\rangle \left\langle J_{a}\right\Vert j^{e}_{1}\left\Vert J_{b}\right\rangle \\
 &  & \times [(-1)^{j^{e}_{1}}c_{k0}(j^{e}_{1}j^{g}_{1})+(-1)^{j^{g}_{1}}c_{k0}(j^{g}_{1}j^{e}_{1})]c^{\star }_{k0}(j_{2}j_{2})\\
 &  & \times w(J_{b}J_{b}j^{e}_{1}j^{g}_{1},kJ_{a})w(J_{b}J_{b}j_{2}j_{2},kJ_{c})P_{k}(cos\theta ).
\end{array}
\end{equation}
 In \( \overline{\varepsilon_{e}(\theta )} \) ,\( \overline{\varepsilon_{f}(\theta )} \)
and \( \overline{\varepsilon_{g}(\theta )} \) \( k \) is the smallest even
integer of the sets \( \left\{2j^{e}_{1},2j_{2},2J_{b}\right\} \), \(\left\{ 2j^{f}_{1},2j_{2},2J_{b}\right\} \)
and \( \left\{2j^{g}_{1},2j_{2},2J_{b} \right\} \) respectively. In \( \overline{\varepsilon_{ef}(\theta )} \),
\( \overline{\varepsilon_{fg}(\theta )} \) and \( \overline{\varepsilon_{ge}(\theta )} \)
\( k \) is the smallest even integer (\( k\neq 0) \) of the sets \( \left\{j^{e}_{1}+j^{f}_{1},2j_{2},2J_{b}\right\} \),
\( \left\{j^{f}_{1}+j^{g}_{1},2j_{2},2J_{b}\right\} \) and \( \left\{j^{e}_{1}+j^{g}_{1},2j_{2},2J_{b}\right\} \)
respectively. \\
We can now write

\begin{equation}
\label{res-6}
\overline{\varepsilon (\theta )}=\left| \left\langle J_{c}\right\Vert j_{2}\left\Vert J_{b}\right\rangle \right| ^{2}\left[ \omega_{e}(\theta )+\delta _{1}^{2}\omega_{f}(\theta )+\delta _{2}^{2}\omega_{g}(\theta )+\delta _{1}\omega_{ef}(\theta )+\delta _{2}\omega_{fg}(\theta )+\delta _{1}\delta _{2}\omega_{ge}(\theta )\right] ,
\end{equation}
 where

\begin{equation}
\label{res-7}
\delta _{1}=\frac{\left\langle J_{a}\right\Vert j^{f}_{1}\left\Vert J_{b}\right\rangle }{\left\langle J_{a}\right\Vert j^{e}_{1}\left\Vert J_{b}\right\rangle }\, and\, \delta _{2}=\frac{\left\langle J_{a}\right\Vert j^{g}_{1}\left\Vert J_{b}\right\rangle }{\left\langle J_{a}\right\Vert j^{e}_{1}\left\Vert J_{b}\right\rangle }.
\end{equation}
 Here

\begin{equation}
\label{res-8}
\begin{array}{ccc}
\omega_{e}(\theta ) & = & \sum _{k}z_{k}(1)z_{k}(2)(-1)^{j^{e}_{1}+j_{2}}c_{k0}(j^{e}_{1}j^{e}_{1})c^{\star }_{k0}(j_{2}j_{2})\\
 &  & \times w(J_{b}J_{b}j^{e}_{1}j^{e}_{1},kJ_{a})w(J_{b}J_{b}j_{2}j_{2},kJ_{c})P_{k}(cos\theta ),
\end{array}
\end{equation}
 \( \omega_{f}(\theta ) \) and \( \omega_{g}(\theta ) \) have the same form
with \( j^{e}_{1}\longrightarrow j^{f}_{1} \) and \( j^{e}_{1}\longrightarrow j^{g}_{1} \)
respectively. \\
And

\begin{equation}
\label{res-9}
\begin{array}{ccc}
\omega_{ef}(\theta ) & = & \sum z_{k}(1)z_{k}(2)(-1)^{j_{2}}[(-1)^{j^{e}_{1}}c_{k0}(j^{e}_{1}j^{f}_{1})+(-1)^{j^{f}_{1}}c_{k0}(j^{f}_{1}j^{e}_{1})]c^{\star }_{k0}(j_{2}j_{2})\\
 &  & \times w(J_{b}J_{b}j^{e}_{1}j^{f}_{1},kJ_{a})w(J_{b}J_{b}j_{2}j_{2},kJ_{c})P_{k}(cos\theta ),
\end{array}
\end{equation}

\begin{equation}
\label{res-10}
\begin{array}{ccc}
\omega_{fg}(\theta ) & = & \sum _{k}z_{k}(1)z_{k}(2)(-1)^{j_{2}}[(-1)^{j^{f}_{1}}c_{k0}(j^{f}_{1}j^{g}_{1})+(-1)^{j^{g}_{1}}c_{k0}(j^{g}_{1}j^{f}_{1})]c^{\star }_{k0}(j_{2}j_{2})\\
 &  & \times w(J_{b}J_{b}j^{g}_{1}j^{f}_{1},kJ_{a})w(J_{b}J_{b}j_{2}j_{2},kJ_{c})P_{k}(cos\theta )
\end{array}
\end{equation}
 and

\begin{equation}
\label{res-11}
\begin{array}{ccc}
\omega_{ge}(\theta ) & = & \sum _{k}z_{k}(1)z_{k}(2)(-1)^{j_{2}}[(-1)^{j^{e}_{1}}c_{k0}(j^{e}_{1}j^{g}_{1})+(-1)^{j^{g}_{1}}c_{k0}(j^{g}_{1}j^{e}_{1})]c^{\star }_{k0}(j_{2}j_{2})\\
 &  & \times w(J_{b}J_{b}j^{e}_{1}j^{g}_{1},kJ_{a})w(J_{b}J_{b}j_{2}j_{2},kJ_{c})P_{k}(cos\theta ).
\end{array}
\end{equation}
 The parameters \( \delta _{1} \) and \( \delta _{2} \) can be determined uniquely
by comparison with experiment{[}\ref{goldfarb}{]}. \\
Writing out the expression (\ref{res-6}) in terms of Legendre polynomials,
we get

\begin{equation}
\label{res-12}
\overline{\varepsilon (\theta )}=\aleph \left[ a_{0}+a_{2}P_{2}(cos\theta )+a_{4}P_{4}(cos\theta )\right] .
\end{equation}
 Here 

\[
\aleph =\left| \left\langle J_{c}\right\Vert j_{2}\left\Vert J_{b}\right\rangle \right| ^{2}\left| \left\langle J_{a}\right\Vert j^{e}_{1}\left\Vert J_{b}\right\rangle \right| ^{2},\]
 and the coefficients are
\( a_{0}=z_{0}(1)z_{0}(2)(1+\delta ^{2}_{1}+\delta ^{2}_{2}) \) , \( a_{2}=z_{2}(1)z_{2}(2)(1.0204+0.751\delta ^{2}_{1}+0.8\delta _{2}^{2}+0.269\delta _{2}+0.311\delta _{1}+0.733\delta _{1}\delta _{2}) \)
and \( a_{4}=z_{4}(1)z_{4}(2)(0.5510-0.122\delta ^{2}_{1}) \). \\
This gives,

\begin{equation}
\label{res-13}
\overline{\varepsilon }=\aleph ^{\prime }\left[ 1+b_{2}P_{2}(cos\theta )+b_{4}P_{4}(cos\theta )\right] ,
\end{equation}
 where \( b_{2}=\frac{a_{2}}{a_{0}}=0.760 \), \( b_{4}=\frac{a_{4}}{a_{0}}=0.042 \)
and \( \aleph ^{\prime }=\aleph a_{0}=\left| \left\langle J_{c}\right\Vert j_{2}\left\Vert J_{b}\right\rangle \right| ^{2}\left| \left\langle J_{a}\right\Vert j^{e}_{1}\left\Vert J_{b}\right\rangle \right| ^{2}a_{0} \).\\
We now define the angular correlation function for this case of channel mixing by writing

\begin{equation}
\label{res-14}
W(\theta )=\left[ 1+b_{2}P_{2}(cos\theta )+b_{4}P_{4}(cos\theta )\right] .
\end{equation}
 Figure 1 gives the results of comparison between our theoretical values and
the experimental values{[}\ref{schmidt}{]}. It will be seen that our polar
plot of \( W(\theta ) \) agrees quite closely with that given by K\( \ddot{a} \)mmerling
and Schmidt {[}\ref{schmidt}{]} except for a small difference in the region
around \( \theta =0 \) degree and \(\theta =180\) degree.

By using scattering theory with the appropriate boundary conditions, it is possible to 
obtain \( \delta _{1} \) and \( \delta _{2} \) without recourse to the experimental
curves. This in its turn should give not only \( a_{0} \), but also \( b_{2} \) and
\( b_{4} \). This is the way our theory can be used to predict the value of
the angular correlation function at any required angle. This would make our
theory autonomous. In the absence of multichannel interaction, \(\delta_{1}=\delta_{2}=0 \), Eq.(\ref{res-14}) reduces to Eq.(\ref{28}).

The expectation value \(\overline{\varepsilon }\) turns out to be the product of a normalization
factor depending on the reduced matrix elements and an angular factor. The simplicity
of the latter is very striking. It seems as if the dynamical calculation
involving the radial matrix elements is redundant. Of course
this is not true, because to extract dynamical properties of the system such
as triply differential cross sections (TDCS) we have to play with the normalizing
factor. However, the very simplicity of the result hinges on the factorization
of the problem into a dynamical part and an geometrical part depending on \( \theta  \).
This comes from the use of Wigner-Eckart theorem, which is a consequence
of the fact that we are dealing with the matrix elements of tensor operators.

Acknowledgments: One of the authors (DC) is deeply grateful to Prof. Volker Schmidt for introducing
him to the subject of double photoionization.

The other author (CS) would like to acknowledge the support provided by the
University Grants Commission of India in the form of a junior research fellowship.

\section{Appendix A }

The following is a description of the notations used in this paper. 

\begin{description}
\item [\( \mathbf{A} \)]: neutral rare gas atom which undergoes double photoionization.
\item [\( e^{-}_{1} \)]: photo-electron.
\item [\( e_{2}^{-} \)]: Auger electron.
\item [\( \mathbf{J_{a}},\mathbf{J_{b}},\mathbf{J_{c}} \)]: total angular momentum vectors of initial, intermediate and
final states of the rare gas atom respectively.
\item [\( J_{a},J_{b},J_{c} \)]: total angular momentum quantum numbers of initial, intermediate and
final states of the rare gas atom respectively.
\item [\( M_{a},M_{b},M_{c} \)]: projection quantum numbers for the initial, intermediate
and final states of the rare gas atoms respectively. 
\item [\( \alpha _{a},\alpha _{b},\alpha _{c} \)]: other quantum numbers corresponding
to initial, intermediate and final state of the rare gas atoms respectively.
\item [\( \mathbf{j_{1}},\mathbf{j_{2}} \)]: total angular momentum vectors of the photo- and Auger electrons
respectively.
\item [\( j_{1},j_{2} \)]: total angular momentum quantum numbers of the photo- and Auger electrons
respectively.
\item [\( \mu _{1},\mu _{2} \)]: projection quantum numbers corresponding to \( j_{1} \)
and \( j_{2} \) respectively.
\item [\( l_{1},l_{2} \)]: orbital angular momentum quantum numbers of the photo- and Auger electrons
respectively.
\item [\( s_{1},s_{2} \)]: spin quantum numbers of the photo- and Auger electrons respectively.
\item [\( \sigma _{1},\sigma _{2} \)]: spin projections of the photo- and Auger electrons
respectively.
\item [\( \rho  \)]: density operator.
\item [\( \rho _{k_{i}\kappa _{i}} \)]: density or statistical tensor of rank \( k_{i} \) with
\( (2k_{i}+1) \) components, \(\kappa_{i}\) being the projection quantum number denoting one component.
\item [\( \varepsilon  \)]: efficiency operator.
\item [\( \varepsilon _{k_{i}\kappa _{i}} \)]: efficiency tensor of rank \( k_{i} \)
with \( (2k_{i}+1) \) components, \(\kappa_{i}\) being the projection quantum number denoting one component.
\item [\( \epsilon  \)]: probability of detection of any event.
\item [\(\left\langle J_{i}\right\Vert j \left\Vert J_{f}\right\rangle \)]: a typical reduced matrix element.
\item [\( \overline{\varepsilon } \)]: expectation value of the efficiency operator.
\item [\( \Omega _{i} \)]: angular direction of the \( i \) th electron.
\item [\( w \)]: Racah coefficients.
\item [\( C^{j_{1}j_{2}j}_{m_{1}m_{2}m} \)]: Clebsch-Gordan Coefficient obeying the
triangle rule \( \Delta (j_{1}j_{2}j) \) and \( m_{1}+m_{2}=m \).
\item [\( \Im  \)]: summation over all non-observed properties of the atom and the
electrons, e.g. spin states and characteristics of the detecting equipment.
\item [\( c_{k\kappa } \)]: Attenuation factors corresponding to the state of polarization.
\item [\( z_{k}(1) \)]: attenuation factor due to the finite size of the detector which
detects the photo-electron.
\item [\( z_{k}(2) \)]: attenuation factor due to the finite size of the detector which
detects the Auger electron.
\item [\(\epsilon_{p}\)]: energy of the photo-electron.
\item [\(\epsilon_{A}\)]: energy of the Auger electron.

All primed quantum numbers denote virtual states.
\end{description}

\section{References}

\begin{enumerate}
\item \label{hamilton}D.R.Hamilton, Phys. Rev. \textbf{58}, 122(1940).
\item \label{gardner}J.W.Gardner, Proc. of the Phys. Soc.` \textbf{A62},763-779(1949).\\
-Proc. of the Phys. Soc. \textbf{A64},238-249(1951). \\
-Proc. of the Phys. Soc. \textbf{A64},1136-1138(1951). 
\item \label{racah}G.Racah, Phys. Rev. \textbf{84} , 910(1951).
\item \label{biedenharn}L.C. Biedenharn and M.E.Rose, Rev. of Mod. Phys. \textbf{25},
729(1953).
\item \label{coester}F. Coester and J. M. Jauch, Helv. Phys. Acta \textbf{26}, 3(1953). 
\item \label{goldfarb}S.Devons and L.J.B.Goldfarb, Angular Correlations, Encyclopedia
of Physics, Vol.42, 384(Edited by S.Fl\( \ddot{u} \)gge, Springer, Heidelberg, 1957).
\item \label{blum}K. Blum, Density Matrix Theory and Applications (Plenum Press, New York, 1981).
\item \label{terharr}D. ter Haar, Elements of Statistical Mechanics 150 (Holt, Reinhart
and Winston, New York, 1960).
\item \label{our1}In reference {[}\ref{coester}{]} Coester and Jauch defined the
angular correlation function for two successive nuclear radiations as the trace
of \( \varepsilon \rho  \) which actually represents the expectation value
of the efficiency operator \( \varepsilon  \). Other authors {[}\ref{ferguson}{]}
have also used that definition. However, the trace of \( \varepsilon \rho  \)
happens to contain a dynamical factor with dimension \( (energy)^{4} \) multiplying
a function of the angle \( \theta  \) between the successively emitted radiations.
Our definition of the angular correlation function \( W(\theta ) \) drops this
factor 
making it a kinematical quantity, as it should be.
\item \label{rose}M.E.Rose, Phys.Rev. \textbf{91}, 610(1953). 
\item \label{we}D.Chattarji and C.Sur, J. El. Spec. and  Rel. Ph. \textbf{114-116},
153(2001). 
\item \label{selles1}P. Selles et al, J.Phys.B \textbf{20}, 5183(1987).
\item \label{selles2}A.Huetz et al, J.Phys.B \textbf{24}, 1917(1991).
\item \label{schmidt}B.K\( \ddot{a} \)mmerling and V. Schmidt, J.Phys. B \textbf{26},
1141-1161(1991).
\item \label{fano}U.Fano, Phys. Rev. \textbf{90}, 577(1953).
\item \label{ferguson}A.J.Ferguson, Angular Correlation Methods in Gamma-ray Spectroscopy, 16
(North-Holland, Amsterdam, 1965).
\item \label{edmonds}A.R.Edmonds, Angular Momentum in Quantum Mechanics, 100(Princeton Univ.
Press, 1957). 
\item \label{dc}D.Chattarji, The Theory of Auger Transitions (Academic Press, London, 1976).
\end{enumerate}
\vspace{0.5001cm}
{\par\centering {\includegraphics{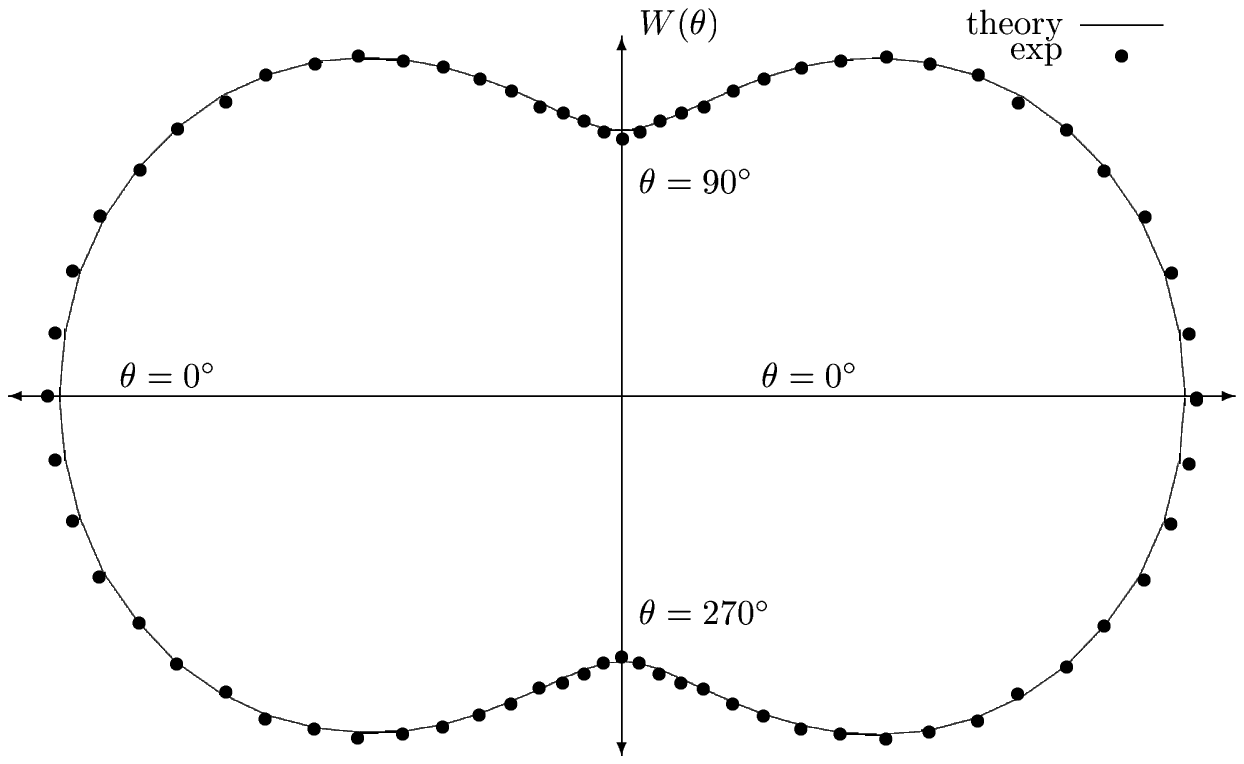}} \par}
\vspace{0.5001cm}

Fig 1: A polar plot of our angular correlation function for DPI in Xenon is
compared with the experimental polar plot given by K\( \ddot{a} \)mmerling
and Schmidt{[}\ref{schmidt}{]}.

\end{document}